# A Formal Comparison of Visual Web Wrapper Generators[*]

Georg Gottlob and Christoph Koch
Database and Artificial Intelligence Group
Technische Universität Wien
A-1040 Vienna, Austria
Email: {gottlob, koch}@dbai.tuwien.ac.at.

**Abstract**

We study the core fragment of the Elog wrapping language used in the Lixto system (a visual wrapper generator) and formally compare Elog to other wrapping languages proposed in the literature.

## 1 Important Note

The long version of the work first presented in the extended abstract [10], which appeared in *Proc. 21st ACM SIGMOD-SIGACT-SIGART Symposium on Principles of Database Systems (PODS 2002)*, Madison, Wisconsin, ACM Press, New York, USA, pp. 17 – 28, was split into two parts for journal publication. The first part [11] studies the expressive power of monadic datalog over trees and establishes the connection to the monadic fragment of the visual wrapper language Elog. The second part, subject of this paper, studies and compares Elog to other practical visual wrapper languages.

Currently, this paper should be understood as an addendum to [11] and cannot be read as a standalone paper. We refer the reader to [11] for all notions of which the definitions are missing.

In detail, this document extends the treatment of [11] by the following material:

- The capability of producing a hierarchically structured result is essential to tree wrapping. We define the language $\text{Elog}_2^*$ in order to be able to make the creation of complex nested structures explicit. $\text{Elog}_2^*$ is basically obtained by enhancing $\text{Elog}^-$ with binary predicates in a restricted form, which allow to represent hierarchical dependencies between selected nodes in the fixpoint computation of an $\text{Elog}^-$ program. $\text{Elog}_2^*$ is an actual fragment of the wrapping language Elog used internally in the Lixto system [7], a commercial visual wrapper generator.

- We take a closer look at two other tree-based approaches to wrapping HTML documents. The first is the language of regular path queries (e.g., [1, 2]) with nesting. Regular path queries are considered essential to Web query languages [1], and by extending the language of regular path queries by capabilities for producing nested output (and for restricting queries by additional conditions), one obtains a useful wrapping language. We show that this formalism is strictly less expressive than $\text{Elog}_2^*$.

[*]This research was supported by the Austrian Science Fund (FWF) under project No. Z29-N04 and the GAMES Network of Excellence of the European Union. A part of the work was done while the second author was visiting the Laboratory for Foundations of Computer Science of the University of Edinburgh and was sponsored by an Erwin Schrödinger scholarship of the FWF.

- The second formalism that we compare to $\text{Elog}_2^*$ is HEL [18], the wrapping language of the commercially available W4F framework, which is the only tree-based wrapping formalism besides Elog of which a formal specification has been published. Again, we are able to show that HEL is strictly less expressive than $\text{Elog}_2^*$.

## 2 Binary Pattern Predicates and Paths with the Kleene Star and Ranges: $\text{Elog}_2^*$

In this section, we step out of our framework of unary information extraction functions. We enhance $\text{Elog}^-$ by a *limited* form of binary pattern predicates, which allow to explicitly represent the parent-child relationship of the tree computed as a result of the wrapping process, but not more than that. This approach to wrapping is basically a mild generalization of our wrapping framework based on unary information extraction functions. The syntax of the full Elog language employs binary pattern predicates in precisely the same way as shown below. The subtle increase in expressive power will be needed in Section 3 when we compare Elog with other practical wrapping languages. A further feature that we will need in Section 3 will be a way of specifying a path using a regular expression with the Kleene star and a "range". We will call the new language obtained $\text{Elog}_2^*$.

We mildly generalize the predicate $\text{subelem}_\pi$ of Definition 6.1 in [11] to support arbitrary regular expressions $\pi$ over $\Sigma$ (notably, including the Kleene star). Again, $\text{subelem}_\pi(v_0, v)$ is true if node $v$ is reachable from $v_0$ through a downward path labeled with a word of the regular language defined by $\pi$.

A range $\rho$ defines, given an integer $k$, a function that maps each $1 \leq i \leq k$ to either 0 or 1. Given a word $w = w_1 \cdots w_k$, $\rho$ *selects* those $w_i$ that are mapped to 1. A range applies to a set of nodes $S$ (written as $S[\rho]$) as follows. Let $v_1 \cdots v_k$ be the sequence of nodes in $S$ arranged in document order. Then, $S[\rho]$ is the set of precisely those nodes $v_i$ for which $i$ is mapped to 1.

**Definition 2.1** Let $\pi$ be a regular expression over $\Sigma$ and let $\rho$ be a regular expression in the normal form of Proposition 4.13 in [11] which defines a regular word language of density one over the alphabet $\{0, 1\}$. The binary relation $\text{subelem}_{\pi,\rho}$ is defined as the set of all pairs of nodes $\langle v, v' \rangle$ such that $v' \in S[\rho]$, where $S$ is the set of all nodes $v_0$ with $\text{subelem}_\pi(v, v_0)$. □

The normal form of Proposition 4.13 in [11] is a convenient syntax for specifying regular word languages of density one, which in turn allow to elegantly assign a unique word over alphabet $\{0, 1\}$ to a sequence of known length. Note, however, that throughout the remainder of the paper, in all languages that we will discuss, only much weaker forms of ranges will be required that can always be easily encoded as regular expressions of this normal form. For example, a range of the form "$i$-th to $j$-th node" (where $i$ and $j$ are constant) can be specified by a regular expression

$$0^{j-1}.1^{j-i+1}.0^*.$$

**Lemma 2.2** *The predicate $\text{subelem}_{\pi,\rho}$ is definable in MSO over $\tau_{ur}$.*

**Proof.** From the proof of Lemma 5.9 in [11], it is obvious how to define a monadic datalog program $\mathcal{P}_\pi$ which defines a predicate $S$ for the set of all nodes reachable from a node $x$ distinguished by a special predicate. From this we obtain an MSO formula $\varphi_\pi(x, S)$ with the obvious meaning using Proposition 3.3 in [11].

Let $\rho$ be the range definition, as a regular expression over the alphabet $\{0, 1\}$ in the normal form of Proposition 4.13 in [11]. We define an MSO formula $\varphi_\rho(S, Y)$ which is true if there is a word $w$ of length $|S|$ in the language $L(\rho)$ and $Y$ is the set of nodes in $S$ that, when traversed in document order, are at a position which is occupied by a "1" in $w$.

Let $\mathcal{P}_\rho$ be the program shown in the construction for down transitions of the proof of Theorem 4.14 in [11], with a few modifications. Rather than on a list of siblings, we try to match $\rho$ with the set $S$ put into document order. Thus, we have to replace occurrences of "firstchild" and "nextsibling" with analogous relations for navigating the document order $\prec$. For example, an atom nextsibling$(x, y)$ is replaced by $\psi_\prec(x, y)$ (using an input relation $S$), where $\psi_\prec$ is defined in MSO as

$$\psi_\prec(x, y, S) := S(x) \wedge S(y) \wedge x \prec y \wedge (\not\exists z)\, S(z) \wedge x \prec z \wedge z \prec y.$$

That the document-order relation $\prec$ itself is MSO-definable is clear from its definition as a caterpillar expression in Example 2.5 in [11], from Lemma 5.9 in [11], and Proposition 3.3 in [11].

In the down-transition construction from the proof of Theorem 4.14 in [11], the goal is to assign (state assignment) predicates that are actually the symbols of the regular language to be matched. In the same way, the unary query that we are interested in is the predicate "1" defined by program $\mathcal{P}_\rho$. The formula $\varphi_\rho$ such that, given set $S$, $\varphi_\rho(S, Y)$ is true iff $Y$ is the set of all nodes assigned "1" by $\mathcal{P}_\rho$ is obtained from $\mathcal{P}_\rho$ as described in the proof of Proposition 3.3 in [11].

Now, it is easy to see that

$$\text{subelem}_{\pi,\rho}(x, y) := (\forall S)(\forall Y) \left( \varphi_\pi(x, S) \wedge \varphi_\rho(S, Y) \right) \to y \in Y$$

indeed defines the desired relation. $\square$

**Remark 2.3** The previous proof makes it easy to extend the formalism to support also the matching of the range backward (using the reverse document order relation $\succ$ rather than $\prec$), and in particular selecting only the last element matching path $\pi$ (using the range $1.0^*$ and reverse document order). $\square$

Now we are in the position to define the language $\text{Elog}_2^*$.

**Definition 2.4** Let $\text{Elog}_2^*$ be obtained by changing the $\text{Elog}^-$ language as follows. All pattern predicates are now binary and all rules are of the form

$$p(x_0, x) \leftarrow p_0(\_, x_0),\ \text{subelem}_{\pi,\rho}(x_0, x),\ C,\ R.$$

where $\text{subelem}_{\pi,\rho}$ is the predicate of Definition 2.1, $C$ is again a set of condition atoms as for $\text{Elog}^-$ but "contains" is now equivalent to $\text{subelem}_{\pi,\rho}$ (permitting ranges and paths defined by arbitrary regular expressions), and $R$ is a set of pattern atoms of the form $p_i(\_, x_i)$. The underscore is a way of writing a variable not referred to elsewhere in the rule. The predicate "root" is also pro-forma binary and can be substituted as a pattern predicate.[1] $\square$

The meaning of a binary pattern atom $p(v_0, v)$ is that node $v$ is assigned predicate $p$ and the inference was started from a parent pattern at node $v_0$. We define *unary queries* in $\text{Elog}_2^*$ in the natural way, by projecting away the first argument positions of our binary pattern predicates. For instance, a program $\mathcal{P}$ (containing a head predicate $p$) defines the unary query $Q_p := \{x \mid (\exists x_0)\, p(x_0, x) \in \mathcal{T}_\mathcal{P}^\omega\}$ based on $p$.

---
[1] As in $\text{Elog}^-$, we need "root" as a parent pattern "to start with".

**Theorem 2.5** *A unary query is definable in $\text{Elog}_2^*$ iff it is definable in MSO.*

**Proof.** Let $\mathcal{P}$ be an $\text{Elog}_2^*$ program and let $\mathcal{P}'$ be the program obtained from $\mathcal{P}$ by adding a rule
$$p'(x) \leftarrow p(x_0, x).$$
for each pattern predicate appearing in $\mathcal{P}$. It is easy to show by induction on the computation of $\mathcal{T}_{\mathcal{P}'}^\omega$ that replacing each rule
$$p(x_0, x) \leftarrow p_0(\_, x_0), \text{subelem}_\pi(x_0, x), C, p_1(\_, x_1), \ldots, p_n(\_, x_n).$$
of $\mathcal{P}'$ (where $C$ is a set of condition atoms) by
$$p(x_0, x) \leftarrow p'_0(x_0), \text{subelem}_\pi(x_0, x), C, p'_1(x_1), \ldots, p'_n(x_n).$$
does not change the meaning of the program. But then, if we only want to compute the unary versions of the pattern predicates, we can just as well replace the heads $p(x_0, x)$ by $p'(x)$ as well. This leads to a monadic datalog program over $\tau_{ur} \cup \{\text{subelem}_{\pi,\rho}\}$. The theorem now follows immediately from Lemma 2.2 and Proposition 3.3 in [11]. □

The rationale of supporting binary pattern predicates in Elog is to explicitly build the edge relation of an output graph during the wrapping process. The obvious unfolding of this directed graph into a tree is what we consider the result of a wrapper run.

**Definition 2.6** The output language of $\text{Elog}_2^*$ is defined as follows. An $\text{Elog}_2^*$ program $\mathcal{P}$ defines a function mapping each document $t$ to a node-labeled directed graph
$$G = \langle V = \text{dom}_t, \ E = \{\langle v_1, v_2 \rangle \mid p_i(v_1, v_2) \in \mathcal{T}_\mathcal{P}^\omega\}, \ (Q_p)_{p \in P} \rangle$$
where $Q_p = \{v \mid (\exists v') \ p(v', v) \in \mathcal{T}_\mathcal{P}^\omega\}$ and $P$ is the set of pattern predicate names occurring in $\mathcal{P}$. □

The edge relation $E$ constitutes a partial order of the nodes. The graph is acyclic except for loops of the form $\langle v, v \rangle \in E$, which are due to specialization rules that produce such loops. In all other rules with a head $p(x, y)$, $y$ matches only nodes strictly below the nodes matched by $x$ in the tree.

**Lemma 2.7** *Each $\text{Elog}_2^*$ binary pattern predicate is definable in MSO.*

**Proof.** Let $\mathcal{P}$ be an $\text{Elog}_2^*$ program and $r$ be a rule of $\mathcal{P}$ with head $P(x_0, x)$, undistinguished variables $x_{j_1}, \ldots, x_{j_l}$ (i.e., $x_0$ and $x$ are precisely the variables of rule $r$ not contained in this list), and a body that consists of the pattern atoms $P_{i_1}(\_, x_{i_1}), \ldots, P_{i_m}(\_, x_{i_m})$ and the set $B$ of remaining atoms.

We use the representation of $\mathcal{P}$ as a monadic program $\mathcal{P}'$ as described in the proof of Theorem 2.5 to define a formula $\varphi$ such that $\varphi(w_1, \ldots, w_m)$ is true iff there exist nodes $v_1, \ldots, v_m$ such that the atoms $P_{i_1}(v_1, w_1), \ldots, P_{i_m}(v_m, w_m)$ evaluate to true on the input tree. Let
$$\varphi(x_{i_1}, \ldots, x_{i_m}) := (\forall P_1) \cdots (\forall P_n) \ SAT(P_1, \ldots, P_n) \rightarrow \big(P_{i_1}(x_{i_1}) \wedge \cdots \wedge P_{i_m}(x_{i_m})\big)$$
where $SAT$ is obtained from $\mathcal{P}'$ as shown in the proof of Proposition 3.3 in [11]. Clearly,
$$P_r(x_0, x) := (\exists x_{j_1}) \cdots (\exists x_{j_l}) \ \varphi(x_{i_1}, \ldots, x_{i_m}) \wedge B$$
is equivalent to the relation defined by the single rule $r$ in $\mathcal{P}$.

Now let $\mathcal{P}_P \subseteq \mathcal{P}$ be the set of rules in the input program whose head predicate is $P$. The formula
$$P(x_0, x) := \bigvee_{r \in \mathcal{P}_P} P_r(x_0, x)$$
defines the desired relation of pattern predicate $P$. □

**Theorem 2.8** *The relations of $G$ are MSO-definable.*

**Proof.** Let $\mathcal{P}$ be an $\text{Elog}_2^*$ program. The edge relation $E$ is simply the union of the relations defined by each of the pattern predicates in $\mathcal{P}$, i.e. a disjunction of their MSO formulae that we have constructed in the proof of Lemma 2.7. The MSO-definability of the $Q_p$ relations was shown in Theorem 2.5. □

We have seen that $\text{Elog}^-$ has linear-time data complexity (see Theorem 4.1). The fixpoint of an $\text{Elog}_2^*$ program (or an Elog program) and equally the edge relation of the output graph, however, can be of quadratic size.

**Example 2.9** Let $t$ be a tree where all leaves are labeled "l", while all other nodes are labeled "b". The single-rule program

$$p(x_0, x) \leftarrow \text{dom}(\_, x_0), \text{subelem}_{\pi=\bar{l}^*l, \rho=*}(x_0, x).$$

evaluates to a fixpoint of quadratic size at worst. For instance, consider a tree with branch nodes $b_1, \ldots, b_m$ and leaf nodes $l_1, \ldots, l_n$ such that $b_i$ is the parent of $b_{i+1}$ (for $1 \leq i < m$) and $b_m$ is the parent of $l_1, \ldots, l_n$. Here, the binary relation defined by $p$ is $\{\langle b_i, l_j \rangle \mid 1 \leq i \leq m,\ 1 \leq j \leq n\}$. □

**Remark 2.10** Note that in full Elog as currently implemented, a range $[\rho]$ can be put at the end of each rule, such that a rule

$$p(x_0, x) \leftarrow p_0(\_, x_0), \text{subelem}_\pi(x_0, x), C, R\ [\rho].$$

has the meaning that $p(v_0, v)$ is inferred from this rule if $v \in S[\rho]$, where $v' \in S$ iff there is an assignment of the variables in the body of the rule to nodes that renders the body true and $x_0$ is assigned to $v_0$ and $x$ to $v'$. □

## 3 Other Wrapping Languages

In this section, we compare the expressiveness of two further wrapping languages, namely regular path queries with nesting and HEL, the wrapping language of the W4F framework [18], to $\text{Elog}_2^*$.

Other previously proposed wrapping languages were evaluated as well. The majority of previous work is string-based (e.g., TSIMMIS [17], EDITOR [5], FLORID [15], DEByE [12], and Stalker [16]) and artificially restricting these languages in some way to work on trees would not be true to their motivation. Thus, we decided not to include them in this discussion. For some other systems (such as XWrap [14], which is essentially tree-based like W4F or Lixto), no formal specifications have been published which can be made subject to expressiveness evaluations.

Web query languages were also evaluated, but some (e.g., WebSQL [4], WebLOG [13]) are unsuitable for wrapping because they cannot access the structure of Web documents, and others[2] (e.g., WebOQL [3]) are highly expressive query languages that permit data transformations not in the spirit of wrapping.

### 3.1 Regular Path Queries with Nesting (RPN)

The first language we compare to $\text{Elog}_2^*$ is obtained by combining regular path queries [2] with nesting to create complex structures. This new language – which we will call RPN (**R**egular **P**ath queries with **N**esting) – on one hand is simple yet appropriate for defining practical wrappers, and on the other hand serves to prepare some machinery for comparing further wrapping languages later on.

---
[2]For a survey of further Web query languages see [9].

**Definition 3.1** The syntax of *RPN* is defined by the grammar

$$
\begin{array}{rl}
rpn\text{:} & patom\ \text{`.'}\ rpn\ |\ \text{`txt'}\ |\ \text{`('}\ rpn\ \text{`\#'}\ \cdots\ \text{`\#'}\ rpn\ \text{`)'} \\
patom\text{:} & patom_0\ |\ patom_0\ conds \\
patom_0\text{:} & path\ |\ path\ \text{`['}\ range\ \text{`]'} \\
range\text{:} & range_0\ \text{`;'}\ \cdots\ \text{`;'}\ range_0 \\
conds\text{:} & \text{`\{'}\ cond\ \text{`and'}\ \cdots\ \text{`and'}\ cond\ \text{`\}'} \\
cond\text{:} & patom\ \text{`.'}\ cond\ |\ \text{`txt'}\ \text{`='}\ string
\end{array}
$$

where *rpn* is the start production, a "$range_0$" is either '*', $i$, or $i-j$ (where $i$ and $j$ are integers), "path" denotes the regular expressions over HTML tag names, and "string" the set of strings. □

Example 3.3 below shows an RPN wrapper in this syntax.

**Definition 3.2 ((Denotational semantics of RPN))** Let $\pi$ denote a path, $\rho$ a range, $s$ a string, and $v, v'$ tree nodes. Without loss of generality, we assume that every *patom* has a range[3]. The semantics function $\mathbb{E}$ maps, given a tree, each pair of an RPN statement $W$ and a node to a complex object as follows:

$$
\begin{aligned}
\mathbb{E}[\![\pi[\rho]\{Y_1 \text{ and } \ldots \text{ and } Y_n\}.X]\!]v &:= \bigcup\{\mathbb{E}[\![X]\!]v'\ |\ \text{subelem}_{\pi,\rho}(v,v')\text{ is true} \wedge \\
& \qquad\qquad\qquad \mathbb{C}[\![Y_1]\!]v' \wedge \cdots \wedge \mathbb{C}[\![Y_n]\!]v'\} \\
\mathbb{E}[\![X_1\# \ldots \#X_n]\!]v &:= \{\langle \mathbb{E}[\![X_1]\!]v, \ldots, \mathbb{E}[\![X_n]\!]v \rangle\} \\
\mathbb{E}[\![\text{txt}]\!]v &:= \{v.\text{txt}\}
\end{aligned}
$$

Here, $v.\text{txt}$ denotes the string value of a node, the concatenation of all text below node $v$ in the input document. Above we assume that both a range description $\rho$ and $n$ conditions are present in a *patom*, but it is clear how to handle the cases where either one or both are missing.

This definition makes use of the semantics function

$$\mathbb{C}: L(cond) \to \text{dom} \to Boolean$$

for RPN conditions, which we define as follows.

$$
\begin{aligned}
\mathbb{C}[\![\pi[\rho]\{Y_1 \text{ and } \ldots \text{ and } Y_n\}.X]\!]v &:= (\exists v')\ \text{subelem}_{\pi,\rho}(v,v')\text{ is true} \wedge \\
& \qquad\qquad\qquad \mathbb{C}[\![Y_1]\!] \wedge \cdots \wedge \mathbb{C}[\![Y_n]\!] \wedge \mathbb{C}[\![X]\!] \\
\mathbb{C}[\![\text{txt} = s]\!]v &:= \textbf{if } v.\text{txt} = s \textbf{ then } true \textbf{ else } false
\end{aligned}
$$

Given a tree $t$, an RPN statement $X$ evaluates to $\mathbb{E}[\![X]\!]root_t$. □

RPN statements can be strongly typed. It is easy the verify that an RPN statement $W$ evaluates to a complex object of type $\mathbb{T}[\![W]\!]$ on all trees, where

$$
\begin{aligned}
\mathbb{T}[\![patom\ X]\!] &:= \mathbb{T}[\![X]\!] \\
\mathbb{T}[\![(X_1\# \ldots \#X_n)]\!] &:= \{\langle \mathbb{T}[\![X_1]\!], \ldots, \mathbb{T}[\![X_n]\!]\rangle\} \\
\mathbb{T}[\![.\text{txt}]\!] &:= \{String\}
\end{aligned}
$$

for *rpn* statements $X, X_1, \ldots, X_n$.

**Example 3.3** The RPN statement

$$\text{html.body.table.tr}\{\text{td}[0].\text{txt} = \text{``item''}\}.\text{td}[1].\text{txt}$$

---

[3]We can always add a range [*] to a *patom* without a range without changing the semantics.

selects the second entries ("td[1]") of table rows ("table.tr") whose first entries have text value "item". The type of this statement is

$$\mathbb{T}[\![\text{html.body.table.tr}\{\text{td}[0].\text{txt} = \text{``item''}\}.\text{td}[1].\text{txt}]\!] = \{String\}$$

Note in particular that the type is not $\{\{String\}\}$, even though there are two patoms not counting the condition! □

**Remark 3.4** Note that the semantics of paths and conditions in RPN is similar to the semantics of a fragment obtained from XPath [19] by prohibiting most of its function library (and therefore its arithmetic and string manipulation features). The simple RPN wrapper of Example 3.3 is basically equivalent to the XPath query

$$/\text{html}/\text{body}/\text{table}/\text{tr}[\text{td}[1] = \text{''item''}]/\text{td}[2].$$

A path of the form $\cdots //a/\cdots$ in XPath corresponds to $\cdots \_^*.a.\cdots$ in RPN.

The main difference is that while XPath selects nodes of the input tree, RPN extracts text below nodes rather than selecting the nodes themselves. Another significant difference between XPath and RPN is that RPN statements may create complex objects (using the nesting construct) that cannot be built in XPath. □

Next, we will show that each wrapper expressible in RPN is also expressible in $\text{Elog}_2^*$. Clearly, there is a mismatch between the forms of output $\text{Elog}_2^*$ and RPN produce which needs to be discussed first. The former language produces trees while the latter produces complex objects containing records.

In the following, we will require $\text{Elog}_2^*$ programs to be of a special form that allows for a canonical mapping from the binary atoms computed by an $\text{Elog}_2^*$ program to a complex object.

Given an RPN statement $W$, each predicate must be uniquely associated to one set or record entry subterm of the type term $\mathbb{T}[\![W]\!]$.

- For a predicate $p$ that is associated to a distinguished set of $\mathbb{T}[\![W]\!]$, an atom $p(v, w)$ asserts that node $w$ is in a set of the output object uniquely identified by $p$ and $v$.

- For a predicate $p$ that is associated to a distinguished (set-typed) record entry of $\mathbb{T}[\![W]\!]$, an atom $p(v, w)$ asserts that $w$ is an element of a record entry in the output object uniquely identified by $p$ and $v$.

By ordering predicates[4] defining the entries of an RPN record appropriately and mapping two nodes $w_1, w_2$ such that there is a node $v$ and two edges $\langle v, w_1 \rangle, \langle v, w_2 \rangle$ labeled with the same predicate into a common set, we obtain the desired mapping to the complex object model of RPN.

It can be easily argued that the distinction between the complex object model of RPN and the output of an $\text{Elog}_2^*$ program that satisfies the above-designed semantics is only cosmetical, indeed that what we produce is a canonical representation of a complex object data model by binary atoms.

RPN also produces string values, while we have not discussed the form in which a tree node is output in Elog so far. We assume that the output of Elog for a node is the concatenation of all text below the node in the document tree. We also assume that text strings are accessible in the document tree (say, string "text" is represented as path-shaped subtree $t \to e \to x \to t \to \bot$) and can be checked using the predicate $\text{contains}_{\pi,\rho}$. (For instance, we can check whether a node $x$ has string value "text" using $\text{contains}_{t.e.x.t.\bot, 1^*}(x, y)$, where $y$ is a dummy variable.)

---

[4]Note that in the reference implementation of Elog (the Lixto system [7, 8]), an ordering of pattern predicates can be defined such that edges of the tree unfolding of the output graph with a common parent node are ordered by their predicate (rather than by document order).

**Theorem 3.5** *For each wrapper expressible in RPN, there is an equivalent wrapper in $Elog_2^*$.*

**Proof.** Ranges in RPN are regular and can be encoded using the $\text{subelem}_{\pi,\rho}$ and $\text{contains}_{\pi,\rho}$ predicates. Clearly, each RPN range can be easily encoded as an $Elog_2^*$ range. Without loss of generality, let $W$ be an RPN statement in which every *patom* has a range. We create the $Elog_2^*$ program $\mathcal{P} := \mathcal{P}_\mathbb{E}[\![W]\!](\text{root})$ using the function $\mathcal{P}_\mathbb{E}$ which maps each pair of an RPN statement and a "context" predicate to an $Elog_2^*$ program, and which is defined as follows.

$$\mathcal{P}_\mathbb{E}[\![\pi[\rho]\{X_1 \text{ and } \ldots \text{ and } X_n\}.Y]\!](p_0) :=$$
$$\{ p'(x_0, x) \leftarrow p(\_, x_0), \text{subelem}_{\pi,\rho}(x_0, x), r_1(\_, x), \ldots, r_n(\_, x). \} \cup$$
$$\mathcal{P}_\mathbb{C}[\![X_1]\!](r_1) \cup \cdots \cup \mathcal{P}_\mathbb{C}[\![X_n]\!](r_n) \cup \mathcal{P}_\mathbb{E}[\![Y]\!](p'),$$

$$\mathcal{P}_\mathbb{E}[\![(X_1\# \ldots \#X_n)]\!](p) := \mathcal{P}_\mathbb{E}[\![X_1]\!](p) \cup \cdots \cup \mathcal{P}_\mathbb{E}[\![X_n]\!](p),$$
$$\mathcal{P}_\mathbb{E}[\![\text{txt}]\!](p) := \emptyset,$$

where $X_1, \ldots, X_n, Y$ are RPN statements, $n \geq 0$, $\pi$ is a path, $\rho$ is a range, and $p', r_1, \ldots, r_n$ are new predicates.

As an auxiliary function for conditions, we have $\mathcal{P}_\mathbb{C}$, defined as

$\mathcal{P}_\mathbb{C}[\![\pi[\rho]\{X_1 \text{ and } \ldots \text{ and } X_n\}.Y]\!](p) :=$
$\quad \{ p(x_0, x) \leftarrow \text{dom}(x_0, x), \text{contains}_{\pi,\rho}(x, y), r_1(\_, y), \ldots, r_n(\_, y), s(\_, y). \} \cup$
$\quad \mathcal{P}_\mathbb{C}[\![X_1]\!](r_1) \cup \cdots \cup \mathcal{P}_\mathbb{C}[\![X_n]\!](r_n) \cup \mathcal{P}_\mathbb{C}[\![Y]\!](s)$

$$\mathcal{P}_\mathbb{C}[\![\text{txt} = s]\!](p) := \{ p(x_0, x) \leftarrow \text{dom}(x_0, x), \text{contains}_s(x, y). \}$$

where $s$ is a string.

A number of predicates generated in this way may correspond to patoms that are followed by further patoms in the RPN statement $W$ and for which no corresponding set exists in $\mathcal{T}[\![W]\!]$ (see Example 3.3).

The reference implementation of Elog, Lixto, allows to define pattern predicates of a given Elog program as auxiliary. Atoms $p(v_0, v)$ of such predicates are then removed from the result of a wrapper run such that if atom $p'(v, w)$ has also been inferred, we add $p'(v_0, w)$ (closing the "gap" produced by dropping the auxiliary predicate).

It is easy to see that the described mapping produces an $Elog_2^*$ program that, when auxiliary predicates are eliminated in this way, maps to canonically to RPN complex objects. □

**Theorem 3.6** *There is an $Elog_2^*$ wrapper for which no equivalent RPN wrapper exists.*

**Proof.** For trees of depth one, all RPN queries are first-order. We therefore cannot check whether, say, the root node has an even number of children, which we can do in MSO and thus, by Theorem 2.5, in $Elog_2^*$. □

## 3.2 HTML Extraction Language (HEL)

In this section, we compare the expressive power of the HTML Extraction Language (HEL) of the World Wide Web Wrapper Factory (W4F) with the expressiveness of $Elog_2^*$. For an introduction to and a formal specification of HEL see [18].

Defining the semantics of HEL is a tedious task. (The denotational semantics provided in [18] takes nearly nine pages and does not yet cover all features!) Here, we proceed in three stages to cover HEL reasonably well. We will define a fragment of HEL called HEL$^-$ which drops a number of marginal features and introduce a slightly simplified version of it, HEL$^-_{vf}$, which does not use HEL's *index variables*. HEL$^-_{vf}$ has the desirable property that the semantics of HEL$^-$ and HEL$^-_{vf}$ entail a one-to-one relationship between wrappers in the two languages [6]. This variable-free syntax is possible because of the very special and restricted way in which index variables may be used in HEL. For simplicity, we first introduce HEL$^-_{vf}$ and subsequently HEL$^-$. Finally, we discuss the remaining features of HEL.

Let RPN$^-$ be the fragment of RPN obtained by requiring that all *patoms* are restricted to the form $t$ or $\_^*.t$ (we will write the latter as $\to t$ ), where $t$ is a tag, and conditions may not be nested inside conditions.

The language HEL$^-_{vf}$ (that is, variable-free HEL$^-$) differs from RPN$^-$ semantically in that ranges apply only to those nodes for which all given conditions hold (i.e., intuitively, conditions are evaluated "first").

Let $\pi$ be either $.t$ or $\to t$ (where $t$ is a tag), and let $\pi_1 \ldots \pi_m$ be paths without conditions. We denote the HEL$^-_{vf}$ semantics function $\mathbb{H}$ (with ranges) by

$$\mathbb{H}[\![\pi[\rho]\{\pi_1.txt = s_1 \wedge \cdots \wedge \pi_m.txt = s_m\}.X]\!]v :=$$
$$\{\mathbb{H}[\![X]\!]w \mid w \in R_\rho(\mathbb{E}[\![\pi\{\pi_1.txt = s_1 \wedge \cdots \wedge \pi_m.txt = s_m\}]\!]v)\}$$

where $\mathbb{E}$ is the RPN semantics function and $R_\rho(V)$ denotes the set of nodes of $V$ matching the range w.r.t. document order, e.g. for range $i$

$$R_i(V) := \{y_i \mid (\exists y_0) \cdots (\exists y_{i-1}) \ y_0, \ldots, y_i \in V \wedge \neg \exists y_{-1} \in V : y_{-1} \prec y_0 \wedge$$
$$\bigwedge_{0 \leq k < i}(y_k \prec y_{k+1} \wedge \neg \exists y' \in V : y_k \prec y' \prec y_{k+1})\}.$$

This selects the $i+1$-th node of $V$. (In HEL, the index of the first node is 0.)

On the remaining forms of HEL$^-_{vf}$ statements $(X_1 \# \ldots \# X_n)$ and txt, $\mathbb{H}$ is defined analogously to $\mathbb{E}$.

**Theorem 3.7** *(1) For each wrapper expressible in the HEL$^-_{vf}$ language, there is an equivalent wrapper in Elog. (2) There is an Elog$_2^*$ wrapper for which no equivalent HEL$^-_{vf}$ wrapper exists.*

**Proof.** (1) can be shown using essentially the same proof as that of Theorem 3.5, with the difference that we use a feature of Elog (see e.g. [8]) that allows to put ranges on the nodes over which the variable $x$ ranges (relative to $x_0$) and replace rules

$$p'(x_0, x) \leftarrow p(\_, x_0), \text{subelem}_{\pi, \rho}(x_0, x), r_1(\_, x), \ldots, r_n(\_, x).$$

by

$$p'(x_0, x) \leftarrow p(\_, x_0), \text{subelem}_\pi(x_0, x), r_1(\_, x), \ldots, r_n(\_, x) \ [\rho].$$

(2) can be justified by the same argument used previously for showing Theorem 3.6. □

Next we discuss the HEL$^-$ language, a proper fragment of HEL. The syntax of HEL$^-$ is considerably different from that of HEL$^-_{vf}$, using a form of index variables in ranges and a special "where" block at the end of a wrapper statement that collects all of the conditions, similar to database query languages such as SQL. To give a better overview of the language, we provide its full syntax.

**Definition 3.8** The syntax of the language HEL$^-$ is defined by the following grammar.

> *HEL$^-$*:    *cc* | *cc* 'where' *conds*
> *cc*:    *pseq*.txt | *pseq* '(' *cc* '#' $\cdots$ '#' *cc* ')'
> *pseq*:    *patom* (('.'|'$\to$') *patom*)$^*$

| | |
|---|---|
| *patom*: | tag \| tag '[' *vrange* ']' |
| *vrange*: | *range* \| var ':' *range* \| var |
| *conds*: | *cond* 'and' ⋯ 'and' *cond* |
| *cond*: | *pseq*.'txt' = string |

where "var" is a set of index variable names, "int" is the set of integers, "tag" the set of HTML tag names, "string" the set of strings, and *range* is defined as in RPN (see Definition 3.1). □

There are a number of further syntactical conditions that restrict the way in which variables can be used in a wrapper. Each index variable used in a $\text{HEL}^-$ statement occurs *exactly once* in its *cc* construct. Moreover, let $P$ the set of paths that can be constructed by concatenating paths in the cc construct starting from the left and always choosing one element of a record while going to the right. Each cond construct $c$ in the where clause of a wrapper is constrained in that the smallest prefix of $c$ that contains all ranges with index variables has to match a prefix of a path (in terms of both tags and index variables appearing in ranges) in $P$.

For example,

html.body.table(tr[0].td[0].txt # tr[i:*].td[1].txt)
where html.body.table.tr[i].td[0].txt = "item";

is correct HEL, because we can construct the path html.body.table.tr[i:*].td[1].txt while reading the *cc* construct from left to right, and this path and its index variables math the condition. (There is a single index variable $i$ occurring in both paths at the same position, and the prefix html.body.table.tr is the same.)

The semantics of $\text{HEL}^-$ will not be introduced in detail but index variables are simply a tool to relate paths in the first "construction" part of the wrapper (everything up to the where clause) with conditions in the second part.

A $\text{HEL}^-$ wrapper can be easily transformed into $\text{HEL}^-_{vf}$ by simply removing its conditions one by one and merging them into the construction part of the wrapper. Starting from the left, each condition is deleted up to the rightmost of its variables, and the remaining condition is nested into the construction part of the wrapper at the position of that variable. For example, the HEL wrapper shown above can be written as

html.body.table(tr[0].td[0].txt # tr[*]{td[0].txt = "item"}.td[1].txt);

in $\text{HEL}^-_{vf}$.

**Proposition 3.9 ([6])** *A wrapper is expressible in $\text{HEL}^-$ iff it is expressible in $\text{HEL}^-_{vf}$.*

Therefore, $\text{HEL}^-$ inherits the expressiveness results of Theorem 3.7.

$\text{HEL}^-$ is the fragment of HEL obtained by taking HEL without string extraction using *match* and *split* expressions (although we support strings in conditions as essential to the philosophy of HEL) and without the getNumberOf and getAttr functions. Note that this is done to compare HEL in our framework based on the language $\text{Elog}_2^*$. Full Elog again supports string extraction in the way HEL does. Using the getNumberOf function of HEL, one may require that the number of nodes (in the document tree) reachable through a given path starting from some node is equal to some constant number, which is easy to define in MSO. The getAttr function of HEL extracts HTML attributes, which we manage as tree nodes. In our framework, the function is redundant with those for accessing nodes.

Some HEL statements can be required to be single-valued (i.e., for a statement $W$ relative to node $v$, $\mathbb{E}[\![W]\!]v$ must contain exactly one node). This is in particular true for condition paths, which must always be single-valued. These issues are best handled at runtime (during complex object creation) using an exception handling mechanism as in W4F.

**Remark 3.10** HEL also supports some form of Prolog-like cut "!" with which some conditions can be marked. The cut causes the evaluation of a path to stop if a condition marked with the cut is false. The HEL cut, however, has not been covered in the formal semantics definition of [18] or unambiguously explained elsewhere. Several different meanings are imaginable.

Let us consider one meaning of the cut, where, given a node $v$, we first evaluate the path $\pi$, and then remove all nodes $w$ that either violate a condition or for which there is a different node $w_0$ such that $w_0$ is reachable from $v$ through $\pi$ and $w_0 \prec w$.

We can formally denote the changed semantics of paths with conditions and the cut (but without ranges) by a semantics function $\mathbb{H}_0$ such that

$$\mathbb{H}_0[\![\pi\{\pi_1.txt = s_1 \wedge \cdots \wedge \pi_m.txt = s_m\}]\!]v :=$$
$$\{z \mid \text{subelem}_\pi(v,z) \wedge C(z) \wedge (\forall x)\ (x \preceq z \wedge \text{subelem}_\pi(v,x)) \to C^!(x)\}$$

where $C(v) := \bigwedge_{1 \leq k \leq m} \mathbb{C}[\![\pi_k.txt = s_k]\!]v$ and $C^!(z)$ if for all conditions $\pi_k.txt = s_k$ with the cut, $\mathbb{C}[\![\pi_k.txt = s_k]\!]v$ is true.[5] This semantics function $\mathbb{H}_0$ can easily be integrated into the above-described function $\mathbb{H}$ to cover ranges as well.

This essentially provides us with a definition of the edge relation that determines the complex objects computed by full HEL wrappers in MSO (see the previous section where we have discussed the relationship between such a binary relation and complex objects). It follows that all unary HEL queries (for any reasonable definition of such queries) are definable in $\text{Elog}^-$. □

# Acknowledgments

We thank Fabien Azavant for insightful discussions.

---

[5]Here we assume that conditions $c$ marked with the cut can fail independently of the order in which the conditions appear in the HEL statement. Alternatively, one could e.g. assume that the cut in $c$ applies only if no condition not marked with a cut that appears left of $c$ in the HEL statement fails on a given node.